\documentclass[aps,prl,reprint,showpacs,superscriptaddress,floatfix]{revtex4-1}

\usepackage{amsmath}
\usepackage{scalerel}
\usepackage{xspace} 
\usepackage{graphicx}
\usepackage{natbib}
\usepackage{hyperref}
\usepackage[utf8]{inputenc}
\hypersetup{
	colorlinks   = true, 
	urlcolor     = blue, 
	linkcolor    = black,
	citecolor   = black 
}

\newcommand{\avg}[1]{\langle{#1}\rangle}

\newcommand{\uvec}[1]{\hat{u}_{#1}}

\newcommand{\mm}{\,\ensuremath{{\rm mm}}\xspace} 
\newcommand{\mum}{\,\ensuremath{\mu{\rm m}}\xspace} 
\newcommand{\nm}{\,\ensuremath{{\rm nm}}\xspace}

\newcommand{\ppm}{\,\ensuremath{{\rm ppm}}\xspace}

\newcommand{\MHz}{\,\ensuremath{{\rm MHz}}\xspace}
\newcommand{\kHz}{\,\ensuremath{{\rm kHz}}\xspace}
\newcommand{\Hz}{\,\ensuremath{{\rm Hz}}\xspace}

\newcommand{\mbar}{\,\ensuremath{{\rm mbar}}\xspace}

\newcommand{\muW}{\,\ensuremath{{\rm \mu W}}\xspace}

\newcommand{\W}{\,\ensuremath{{\rm W}}\xspace}

\newcommand{\mK}{\,\ensuremath{{\rm mK}}\xspace}
\newcommand{\K}{\,\ensuremath{{\rm K}}\xspace}

\newcommand{\trbkUnpro}{\raisebox{-1pt}{\scalerel*{\includegraphics{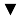}}{B}}}
\newcommand{\dibkUnpro}{\raisebox{-0.5pt}{\scalerel*{\includegraphics{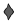}}{B}}}
\newcommand{\cibkUnpro}{\raisebox{-1pt}{\scalerel*{\includegraphics{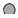}}{B}}}

\newcommand{\trbk}{\protect\trbkUnpro}
\newcommand{\dibk}{\protect\dibkUnpro}
\newcommand{\cibk}{\protect\cibkUnpro}

\begin{document}
	
	\title[]{Cavity-Based 3D Cooling of a Levitated Nanoparticle via Coherent Scattering}

	\author{Dominik~\surname{Windey}}	
	\affiliation{Photonics Laboratory, ETH Z{\"u}rich, 8093 Z{\"u}rich, Switzerland}
	\author{Carlos~\surname{Gonzalez-Ballestero}}
	\affiliation{Institute for Quantum Optics and Quantum Information of the Austrian Academy of Sciences, A-6020 Innsbruck, Austria}
	\affiliation{Institute for Theoretical Physics, University of Innsbruck, A-6020 Innsbruck, Austria.}
    \author{Patrick~\surname{Maurer}}
    \affiliation{Institute for Quantum Optics and Quantum Information of the Austrian Academy of Sciences, A-6020 Innsbruck, Austria}
	\affiliation{Institute for Theoretical Physics, University of Innsbruck, A-6020 Innsbruck, Austria.}
    \author{Lukas~\surname{Novotny}}
    \affiliation{Photonics Laboratory, ETH Z{\"u}rich, 8093 Z{\"u}rich, Switzerland}
    \author{Oriol~\surname{Romero-Isart}}	
    \affiliation{Institute for Quantum Optics and Quantum Information of the Austrian Academy of Sciences, A-6020 Innsbruck, Austria}
	\affiliation{Institute for Theoretical Physics, University of Innsbruck, A-6020 Innsbruck, Austria.}
	\author{Ren\'{e}~\surname{Reimann}}
	\email{rreimann@ethz.ch}
	\affiliation{Photonics Laboratory, ETH Z{\"u}rich, 8093 Z{\"u}rich, Switzerland}

	
	\begin{abstract}
	We experimentally realize cavity cooling of all three translational degrees of motion of a levitated nanoparticle in vacuum. 
	The particle is trapped by a cavity-independent optical tweezer and coherently scatters tweezer light into the blue detuned cavity mode. 
	For vacuum pressures around $10^{-5}\,{\rm mbar}$,  minimal temperatures along the cavity axis in the millikelvin regime are observed.
	Simultaneously, the center-of-mass (c.m.) motion along the other two spatial directions is cooled to minimal temperatures of a few hundred millikelvin. 
	Measuring temperatures and damping rates as the pressure is varied, we find that the cooling efficiencies depend on the particle position within the intracavity standing wave.
	This data and the behavior of the c.m. temperatures as functions of cavity detuning and tweezer power are consistent with a theoretical analysis of the experiment. 
	Experimental limits and opportunities of our approach are outlined.
	\end{abstract}

	\maketitle
	
\textit{Introduction.}---Arthur Ashkin pioneered the use of light to control minute particles. 
His early work on optical tweezers~\cite{Ashkin1970, Ashkin1976} is currently experiencing a renaissance in the modern field of levitated optomechanics. 
This rapidly developing field optically manipulates mesoscopic particles in vacuum to investigate thermodynamics~\cite{Gieseler2018} and rotational dynamics~\cite{Kuhn2017a, Shi2016} on the nanoscale, or---quite practically---pushes the limits of ultrasensitive sensing~\cite{Ranjit2016, Hebestreit2018a, Monteiro2017,Kuhn2017}. 
All of these areas of levitated optomechanics rely on tightest control over the center-of-mass (c.m.) motion of the levitated particle.
The resulting experimental c.m. cooling efforts can be divided into an active and a passive approach. 
For active cooling, the particle c.m. position is measured and---using electronic data processing and subsequent negative feedback---applied back to the oscillator~\cite{Gieseler2012, Li2011b, Tebbenjohanns2018}. 
In contrast, passive cooling is based on the idea of introducing a cavity with a narrow optical resonance, which can be used to lower the particle’s c.m. energy via enhanced anti-Stokes scattering~\cite{Vuletic2000}.

Passive cavity cooling was first applied in atomic systems~\cite{Horak1997, Ye1999, Vuletic2001, McKeever2003, Maunz2004, Nussmann2005, Leibrandt2009, Wolke2012, Hosseini2017}, but has soon been adapted to levitated optomechanics~\cite{RomeroIsartNJP2010,Chang2010,Romero-Isart2011a}. 
There, experiments focused on one-dimensional cavity cooling realized by directly driving the cavity. 
The particle was trapped via an additional intracavity light field~\cite{Kiesel2013}, or via a hybrid electro-optical trap~\cite{Millen2015, Fonseca2016}, achieving minimal temperatures of $\lesssim 0.3\K$ along the cavity axis. 

Going back to Ashkin’s early ideas, in our experiment, we minimize technological complexity and increase the level of control by trapping the particle in an optical tweezer, which---similar to Ref.~\cite{Magrini2018}---is geometrically independent from the cavity. 
However, in contrast to Ref.~\cite{Magrini2018}, our tweezer light is near-resonant to the optical cavity.
Therefore, the particle coherently scatters tweezer light into the cavity, which is slightly blue detuned from the optical frequency of the tweezer trap, leading to position-dependent cavity cooling of \textit{all} motional degrees of freedom.
At certain positions and for low vacuum pressures, we measure temperatures lower than a few hundred millikelvin for all axes. Along the cavity axis, minimal temperatures in the few millikelvin range are reached.

\begin{figure}[h!]
		\centering{\includegraphics[width=0.9\columnwidth]{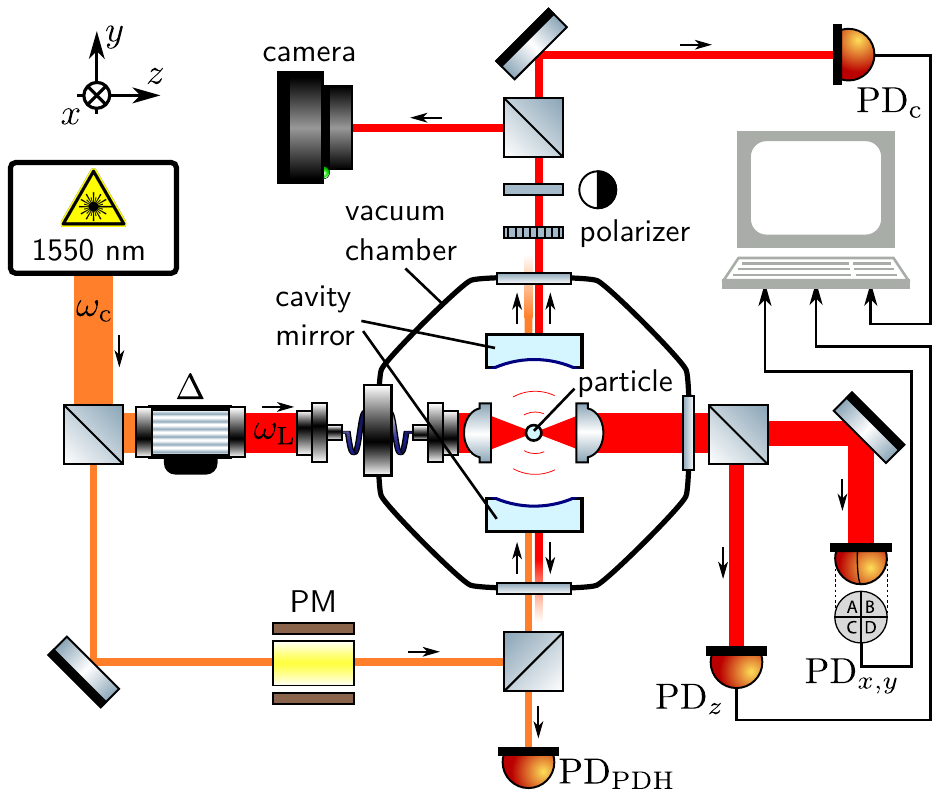}}\\
		\caption{Simplified experimental setup. A single nanoparticle is levitated in an optical tweezer trap and positioned in the mode of an optical cavity. The particle is driven by the trapping light and scatters into the cavity mode and into free space. This scattered light is detected by the cavity photodetector $\rm PD_c$ and by the free space photodetectors ${\rm PD}_z$ and ${\rm PD}_{x,y}$, respectively.}
		\label{Fig.setup}
	\end{figure}

\textit{Experimental setup.}---Our apparatus is shown in Fig.~\ref{Fig.setup}. Laser light at a wavelength $\lambda = 1550.0(5)\nm$ is split into two beams.
The first laser beam with frequency $\omega_{\rm c} = 2\pi c/\lambda$, where $c$ is the speed of light, is modulated by a phase modulator (PM) to generate a lock beam that is coupled into our optical cavity. 
The $z$ polarized lock beam with an optical power of $11(1)\muW$ is back reflected from the cavity and detected with a photodiode ($\rm PD_{PDH}$). From the photodiode signal a Pound-Drever-Hall error signal is derived~\cite{Drever1983}, which we utilize to stabilize the cavity length $L = 6.46(8)\mm$ by means of piezoelectric transducers (not shown). 
The locked cavity with resonance frequency $\omega_{\rm c}$ supports a Gaussian mode with waist $w_0 = 48(5)\mum$. 
The cavity with linewidth $\kappa = 2\pi\times 1.06(8)\MHz$ and finesse $\mathcal{F}=\pi c/(\kappa L) = 22(2)\rm \times 10^3$ is built from two identical mirrors with absorption $A=45(6)\ppm$, transmission $T=99(9)\ppm$ and radius of curvature $\rm ROC = 10.0(1)\mm$.
The second laser beam is frequency shifted by $\Delta$ and used for trapping. 
The resulting light at frequency $\omega_{\rm L} = \omega_{\rm c} - \Delta$ is coupled into the vacuum chamber via a polarization-maintaining optical fiber. 
Inside the chamber, the approximately $x$ polarized light is collimated and sent through a lens with numerical aperture $\rm NA = 0.83$ which forms an optical tweezer trap [focal power $P_{\rm tw}=0.50(5)\W$] for a $\rm SiO_2$ particle with $136\nm$ nominal diameter. 
A second identical lens is rigidly mounted to the first one and collimates the light again, which is then distributed to two free space detectors. 
One of them (${\rm PD}_z$) is measuring the particle c.m. motion along the $z$ direction, while the second, a quadrant photodetector (${\rm PD}_{x,y}$), detects the particle c.m. motion along the $x$ and $y$ direction~\cite{Gieseler2012}. 
Measured c.m. trap frequencies are on the order of $\Omega_{x,y,z} \approx 2\pi \times \{0.12,0.14,0.04\} \MHz$.

Similar to Ref.~\cite{Mestres2015}, the particle in the tweezer trap is positioned in the center of the Gaussian mode of the locked cavity with a three-dimensional (3D) resolution on the $50 \nm$ scale. 
Experimentally, we optimize the coupling of the particle to the cavity mode by scanning the particle position in the $x,z$ plane until we reach a position where the signal on the photodiode $\rm PD_c$ is maximal. 
Additional to the detector $\rm PD_c$, which detects the trapping light the particle scatters into the cavity, we use a camera to assure that the spatial profile of this scattered light is Gaussian. 
In our measurements, the central signal is the 3D c.m. position of the cavity-coupled particle, which we deduce from the voltages of ${\rm PD}_{x,y}$, ${\rm PD}_z$ and $\rm PD_c$. The corresponding time traces are recorded with a sampling rate of $5\MHz$.
	
\textit{Results and discussion.}---For our measurements, which are all taken with the very same single nanoparticle, we follow the calibration and temperature estimation protocols outlined in Ref.~\cite{Hebestreit2018}. 
In short, the calibration relies on the equipartition theorem, while the temperatures are estimated from the areas of power spectral densities calculated from the calibrated signals of ${\rm PD}_{x,y}$ and ${\rm PD}_z$. 
In general, we concentrate on cavity cooling of the c.m. particle motion in all three spatial dimensions. Throughout the manuscript, data corresponding to motion along $x$, $y$ and $z$ are depicted in blue, green, and red, respectively. 
Additionally, up to three different particle positions relative to the standing wave axis ($y$) of the cavity field are considered. 
A particle positioned near the node, steep slope, or antinode of the standing wave is represented by different markers (\cibk, \dibk~and~\trbk). In the experiments, we distinguish those positions by measuring the ${\rm PD_c}$ signal. 
A low, medium, or high signal corresponds to \cibk, \dibk~or~\trbk, respectively.\\

	\begin{figure}[b!]
		\centering{\includegraphics[width=\columnwidth]{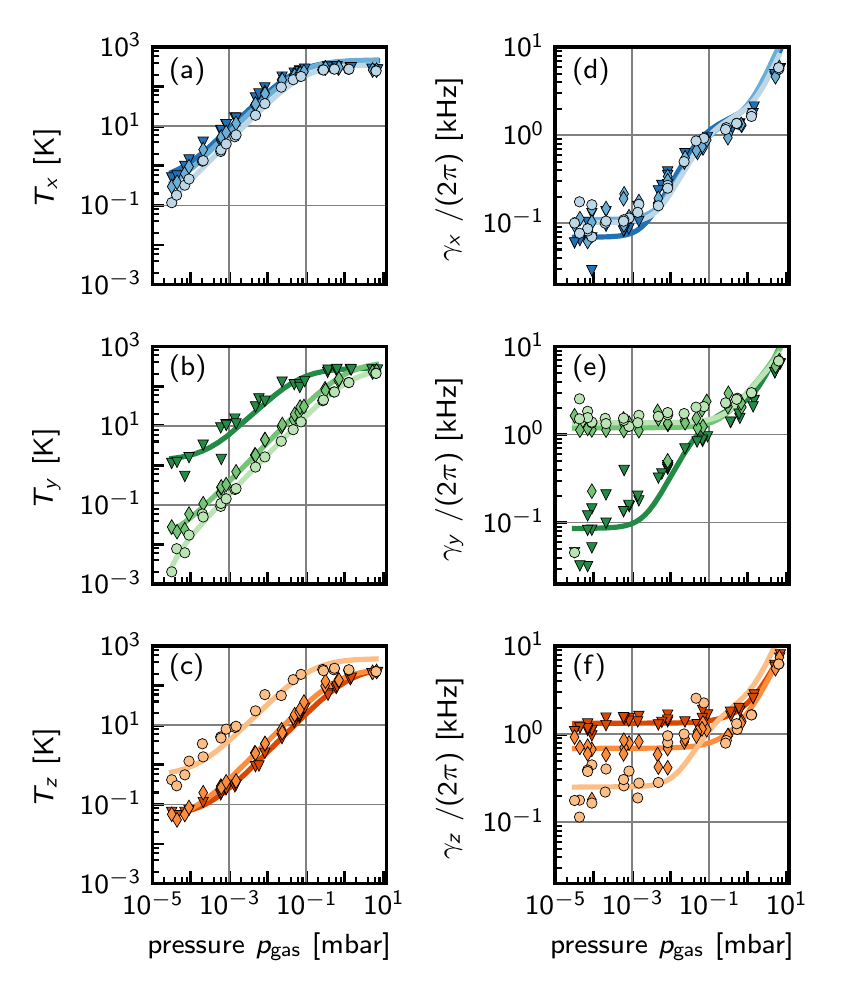}}\\
		\caption{Three-dimensional cavity cooling of an optically levitated nanoparticle by coherent scattering ($\Delta = 2 \pi \times 400 \kHz$). The cooling is compared for a nanoparticle positioned at the node (\cibk), steep slope (\dibk) and anti-node (\trbk) of the cavity standing wave. (a--c) Particle temperatures. (d--f) Particle damping rates. Both, temperatures and damping rates decrease as a function of gas pressure and are position dependent. Solid lines represent a combined fit to a two bath model and a damping rate model.}
		\label{Fig.temp_gamma}
	\end{figure}

In our first measurement, displayed in Fig.~\ref{Fig.temp_gamma}, we study the 3D temperatures $T$ and damping rates $\gamma$ of the particle as a function of gas pressure $p_{\rm gas}$. 
For cavity cooling by coherent scattering, the cavity is blue detuned from the tweezer light ($\Delta = 2 \pi \times 400 \kHz$).
Figures~\ref{Fig.temp_gamma}(a--c) show that the temperatures along all axes decrease, as the pressure and therewith heating due to interaction with room temperature gas molecules are reduced. 
Along $x$ and $y$ we observe lowest temperatures $T_x \approx 100 \mK$  and $T_y \approx 3 \mK$ at the node, limited by interaction with residual gas. 
For $z$, however, we find lowest temperatures $T_z \approx 80 \mK$ at the anti-node, starting to level off around a pressure of $10^{-5}\mbar$. 
The observed position-dependent cooling can be understood by considering the mean optical gradient force $\vec{F}_{\rm grad}$ acting on the particle via the tweezer and the cavity electric field $\vec{E}_{\rm tw}\propto e^{i k z} \uvec{x}$ and $\vec{E}_{\rm c}\propto \cos(k y + \phi)\uvec{x}$, respectively. 
Here $k=2\pi/\lambda$ and $\uvec{i}$ is the unit vector along direction $i$. 
We choose the equilibrium position of the oscillating particle as origin $x,y,z =0$, even though in practice the optical tweezer is shifted and not the cavity. 
This leads to a phase $\phi$ of $\pi/2$ for a particle at the node, and $0$ for a particle at the anti-node of the intracavity field.
Calculating the gradient force via $\vec{F}_{\rm grad}\propto \nabla |\vec{E}_{\rm tw} + \vec{E}_{\rm c}|^2$ one finds that the dominant $\phi$ dependent terms scale with $\sin(\phi)\uvec{y}$ and $\cos(\phi)\uvec{z}$.
Such a position-dependent energy exchange results, together with the fast cavity dissipation, in expected optimal cooling for $\phi=\pi/2$ (particle at node, \cibk) along $y$ and for $\phi =0$ (particle at anti-node, \trbk) along $z$. 
The observed optimal cooling for $\phi=\pi/2$ along the $x$ direction is explained by the tweezer light being not perfectly polarized along $x$. 
This imperfection turns into a feature as one realizes that the main axis of the resulting tweezer trapping potential is not perfectly orthogonal to the cavity axis, which results in cooling along $x$ induced by the same mechanism as described for $y$. A more detailed theoretical description of the observed effects can be found in Ref.~\cite{Ballestero2019}.

In a quantitative approach we fit the data in Figs.~\ref{Fig.temp_gamma}(a--c) to a two bath model with an additional heating rate $\avg{\dot{T}_{\rm noise}}$ possibly arising from optical trap displacement noise~\cite{Gehm1998}. 
In this model, the particle temperature along direction $i$ is given by $T_i = (\gamma_{\rm gas} T_{\rm gas} + \avg{\dot{T}_{\rm noise}})/(\gamma_{\mathrm{gas}} + \gamma_{\mathrm{c},i})$, where  $\gamma_{\mathrm{c},i}$ is the damping rate due to cavity cooling, $\gamma_{\mathrm{gas}}\propto p_{\rm gas}$ is the damping rate due to gas molecule collisions and  $T_{\rm gas} \approx 300 \K$ is the gas temperature.
We find a heating rate of $\avg{\dot{T}_{\rm noise}} = 33(27)\,{\rm K/s}$ which would correspond to an optical trap displacement noise of about $10^{-14}\,{\rm m/\sqrt{Hz}}$~\cite{Gehm1998}.
We observe $\avg{\dot{T}_{\rm noise}}$ to be an order of magnitude higher in $z$ direction compared to $x, y$ which might be connected to the particular response of our experimental system to mechanical noise. A more detailed analysis of the noise in our system is ongoing work.

Figures.~\ref{Fig.temp_gamma}(d--f) display the damping rates $\gamma_i$, which are extracted as the full width at half maximum from the respective power spectral densities. Following Ref.~\cite{Hebestreit2017}, we model the damping rates as $\gamma_i \approx \sqrt{(\gamma_{\mathrm{NL},i})^2 + ( \gamma_{\mathrm{gas}} + \gamma_{\mathrm{c},i} )^2}$ where $\gamma_{\mathrm{NL},i}$ is the broadening of the linewidth due to nonlinearities of the trapping potential. 
At pressure $p_{\rm gas}\gtrsim 1\mbar$,  gas damping dominates. 
Nonlinear broadening, proportional to the particle temperature, is most pronounced in the regime between $10^{-1}$ and $10^{-3}\mbar$ (hump in data) where cavity cooling is not very efficient yet but gas damping has already decreased significantly. 
At sufficiently low pressure and for efficient cavity cooling, the damping rates level off, and $\gamma_i \rightarrow \gamma_{\mathrm{c},i}$ reaching in the best case $2 \pi \times 1.3 \kHz$.
Solid lines in the plot are obtained by simultaneously fitting the two bath model to the data in  Figs.~\ref{Fig.temp_gamma}(a--c) and the damping rate model to the data in Figs.~\ref{Fig.temp_gamma}(d--f).

	\begin{figure}[b!]
		\centering{\includegraphics[width=\columnwidth]{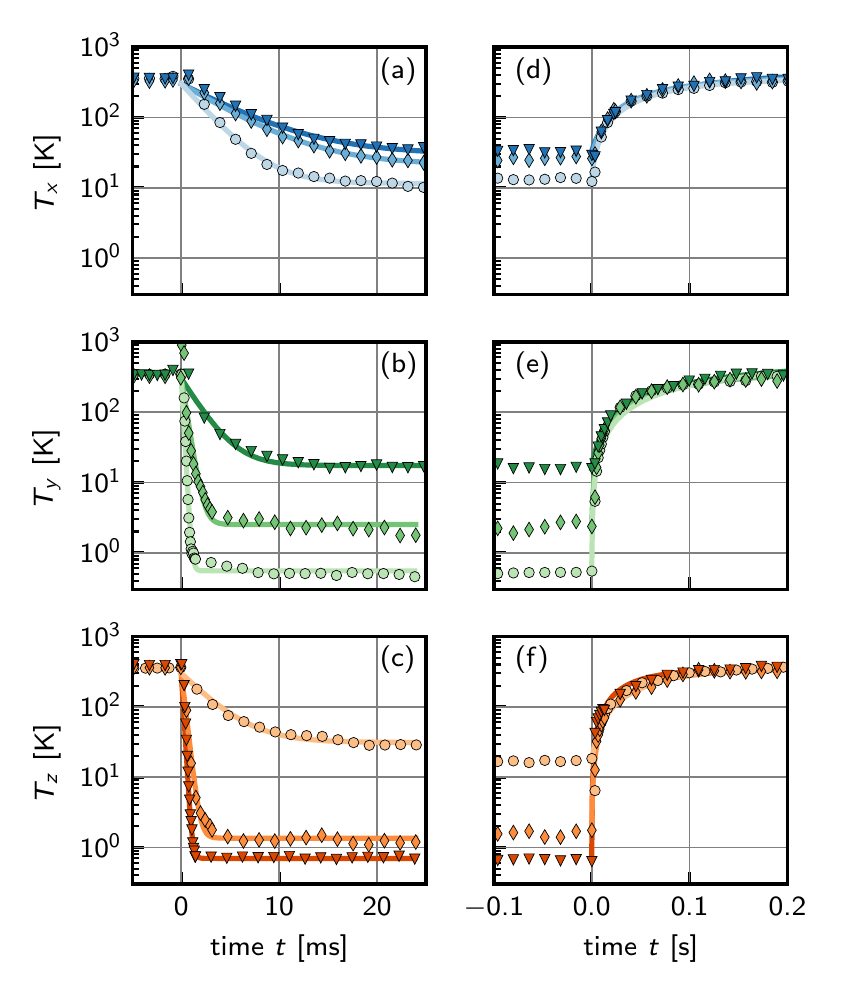}}\\
		\caption{Cavity cooling and reheating time traces of a nanoparticle at $p_{\rm gas}=3(1) \times 10^{-3} \mbar$ averaged over $>150$ realizations. Markers and colors as in Fig.~\ref{Fig.temp_gamma}. (a--c) At $t=0$ cavity cooling is turned on and the decrease of $T_x$, $T_y$ and $T_z$ is shown over time. (d--e) At $t=0$ cavity cooling is turned off and the increase of $T_x$, $T_y$ and $T_z$ is shown over time. Solid lines represent fits of the particle temperatures to a bounded exponential growth model.}
		\label{Fig.relaxation}
	\end{figure}
In our second measurement, see Fig.~\ref{Fig.relaxation}, we study cooling and also heating rates via a time resolving switching method.
Cavity cooling is turned on by switching the detuning $\Delta$ from $2 \pi \times 20 \MHz$ to $2 \pi \times 400 \kHz$ and turned off by switching from $2 \pi \times 400 \kHz$ to $2 \pi \times 20 \MHz$. 
The temperatures at every instant of time are given by the areas of the power spectral densities of short snapshots of the recorded time traces after digital noise filtering. 
Since the nanoparticle occupies a thermal motional state, we analyze the average of more than $150$ realizations.
We measure at $p_{\rm gas}=3(3) \times 10^{-3} \mbar$, as there the nanoparticle motion is mainly damped by cavity backaction, see Figs.~\ref{Fig.temp_gamma}(d--f), and the experiments are not influenced by mechanical drifts of the setup, which occur on the minute timescale. 
The damping rate is therefore equal to the cavity cooling rate, and it can be extracted from monitoring the nanoparticle temperature as a function of time after switching the cavity cooling mechanism on as shown in Figs.~\ref{Fig.relaxation}(a--c). 
We determine the cooling rates by modeling the data as bounded exponential growth $T_i(t) = T_{\infty, i} + \left( T_{0,i} - T_{\infty, i} \right) e^{- \gamma_{\mathrm{c},i} t}$ for $i \in \{x, y, z\}$,  with time $t$, cooling rate $\gamma_{\mathrm{c},i}$, starting equilibrium temperature $T_0$ and end equilibrium temperature $T_\infty$. 
The fitted rates agree better than a factor of five with $\gamma_{\mathrm{c}, i}$ shown in Figs.~\ref{Fig.temp_gamma}(d--f). We attribute the respective deviations to pressure measurement uncertainties and drifts of system parameters (e.g. tweezer power, particle position). 
The reheating data in Figs.~\ref{Fig.relaxation}(d--f) are analyzed analogously, resulting in reheating rates of $\approx 2 \pi \times 2.5(5) \Hz$ 
that coincide for all axes and positions and are limited by gas reheating~\cite{Gieseler2012}. We remark that the trap displacement noise $\avg{\dot{T}_{\rm noise}}$ does not influence the measured rates but only the c.m. temperatures $T_i$~\cite{Ballestero2019}.

\begin{figure}[h!]
		\centering{\includegraphics[width=\columnwidth]{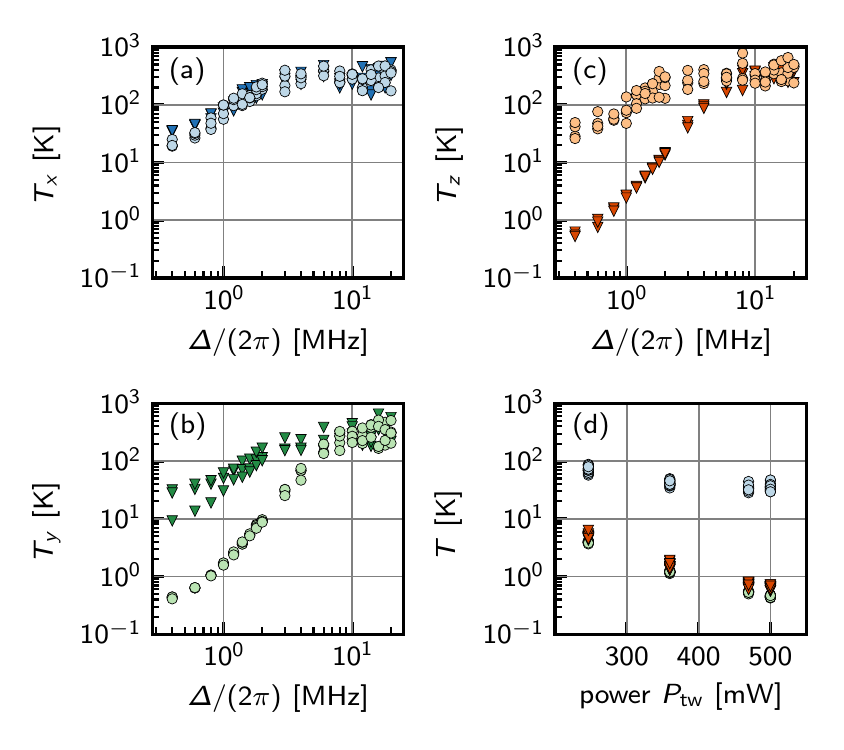}}\\
		\caption{Detuning and power dependence of 3D cavity cooling at $p_{\rm gas} = 3(1) \times 10^{-3} \mbar$. Markers and colors as in Fig.~\ref{Fig.temp_gamma}. (a--c) At tweezer power $P_{\rm tw} = 0.50(5)\W$ the nanoparticle temperatures increase as the cavity detuning $\Delta$ reaches values $\gtrsim\kappa$. (d) Nanoparticle temperatures for best cooling positions of the particle as function of tweezer power at detuning $\Delta = 2 \pi \times 400 \kHz$.}
		\label{Fig.detuning_power}
	\end{figure}

So far, we have used a detuning $\Delta = 2 \pi \times 400 \kHz$ and a tweezer power $P_{\rm tw} = 0.50(5)\W$. Those parameters are identified as ideal for efficient cavity cooling in Fig.~\ref{Fig.detuning_power}. Figures~\ref{Fig.detuning_power}(a--c) show the position dependent particle temperatures as a function of detuning. 
Since the cavity linewidth is large compared to the mechanical frequencies of the particle ($\kappa>\Omega_{x,y,z}$), the optimal detuning $\Delta = 2 \pi \times 400 \kHz$ is approximately the same for all three oscillators. 
For $\Delta \lesssim 2 \pi \times 300 \kHz$ we enter a regime where $g^2 \gtrsim |\Delta|\Omega_{x,y,z}$ and the system becomes dynamically unstable which results in particle loss \cite{Ballestero2019,Kustura2018}. 
Here, $g\gtrsim 2\pi \times 10\kHz$ is the light-enhanced optomechanical coupling rate~\cite{Aspelmeyer2014, Ballestero2019}.

At large detunings, $\Delta\gtrsim2\pi\times 10\MHz$, we observe no influence of the cavity on the particle c.m. temperatures. 
This motivates the chosen detuning of $2\pi\times 20\MHz$ for switching off cavity cooling in the experiments shown in Fig.~\ref{Fig.relaxation}.
The dependence of cooling on the tweezer power is shown in Fig.~\ref{Fig.detuning_power}(d). 
Sweeping the power from $0.24$  to $0.5\W$ results in stronger cavity cooling and lower particle temperatures. 
At powers around $0.5 \W$, however, we observe a saturation of the c.m. temperatures. 
These observations can be explained by noting that the cavity cooling rate scales with the power $P$. 
At higher powers $P$, however, the quadratically growing heating rate~\cite{Gehm1998} $\avg{\dot{T}_{\rm noise}} \propto \Omega_i^4 \propto P^2$ becomes more relevant and could limit the achievable minimal temperatures.
The results of Fig.~\ref{Fig.detuning_power} are well in agreement with a detailed theory, see Ref.~\cite{Ballestero2019}.

\textit{Conclusion and Outlook.}---Our lowest c.m. temperatures $T$ are currently limited by gas pressure and noise, which probably arises from position fluctuations of the trap center~\cite{Gehm1998}. 
After solving those technical problems, the minimal mean phonon number along $y$ would be reduced from its current value on the order of $100$ to $\overline{n}_y\approx\kappa/(4\Omega_y)\approx 2$,~\cite{Aspelmeyer2014}.
As shown, cavity cooling in the fast cavity regime ($\kappa>\Omega_{x,y,z}$) keeps $T_x$ and $T_z$ simultaneously so low, that the trapping potentials along all axes can be considered fully harmonic, and detrimental coupling between the axes, which can lead to heating of $T_y$, is negligible.

To realize c.m. ground state cooling ($n_y\ll 1$) we plan to combine our passive cavity cooling approach with active cooling~\cite{Genes2008, Tebbenjohanns2018}. 
Such a cooling protocol is promising as the scattering into the cavity mode is highly favored due to the Purcell effect~\cite{Kuhn2010, Tanji-Suzuki2011a, Motsch2010}. 
For our system, close to cavity resonance, a fraction $ f= \eta/(\eta+1) \gtrsim 80\%$ of the overall scattered power would be emitted into the cavity, where $\eta = (6\mathcal{F}\lambda^2)/(\pi^3 w_0^2) = 4.4(3)$ is our Purcell factor. 
This high fraction $f$ of the scattered power, containing most of the particle position information along $y$ in a very clean Gaussian cavity mode, can be measured by ${\rm PD_c}$ via a homodyne scheme~\cite{Kiesel2013}. 
This homodyne signal can then be utilized for feedback ground-state cooling~\cite{Gieseler2012, Tebbenjohanns2018}, with the cavity acting as a measurement enhancement device that ensures a high collection efficiency for photons scattered off the particle~\cite{Rodenburg2015}.
Realizing the c.m. motional ground state would introduce levitated optomechanics into the realm of quantum physics~\cite{Aspelmeyer2012, Aspelmeyer2014} and enable the study and usage of mesoscopic nonclassical states of motion~\cite{Romero-Isart2011, Romero-Isart2011a}.

\begin{acknowledgments}
This research was supported by the Swiss National Science Foundation (no.~200021L$\_$169319) and ERC-QMES (no.~338763). R.~R. acknowledges funding from the EU Horizon 2020 program under the Marie Skłodowska-Curie grant agreement no.~702172.  C.~G.~B. acknowledges funding from the EU Horizon 2020 program under the Marie Sk\l{}odowska-Curie grant agreement no.~796725.\\
We thank M.~Frimmer, E.~Hebestreit, R.~Diehl, F.~Tebbenjohanns, F.~van der Laan, A.~Militaru and P.~Mestres for insightful discussions.
\end{acknowledgments}
	
\textit{Note added.}---We have recently become aware of related experimental work by Deli{\'c} et al. in the Aspelmeyer group in Vienna~\cite{Delic2018}.

\bibliographystyle{apsrev4-1}
\bibliography{cavity_cooling_bib_domi_rene}

\end{document}